\documentstyle[prl,aps,floats,epsfig]{revtex}
\newcommand{\be}{\begin{equation}}
\newcommand{\ee}{\end{equation}}
\newcommand{\bees}{\begin{eqnarray}}
\newcommand{\ees}{\end{eqnarray}}
\newcommand{\ra}{\rightarrow}
\newcommand{\gsim}{\stackrel{>}{\sim}}

\newcommand{\ls}{\lambda_{\rm s}}

\begin{document}

\par
\begingroup
%\twocolumn[%

\begin{flushright}
 IFUP-TH 22/97\\
 June 1997\\
gr-qc/9706053
\end{flushright}

{\large\bf\centering\ignorespaces
The fine tuning problem in pre-big-bang inflation
\vskip2.5pt}
{\dimen0=-\prevdepth \advance\dimen0 by23pt
\nointerlineskip \rm\centering
\vrule height\dimen0 width0pt\relax\ignorespaces
Michele Maggiore$^{a,b}$ and Riccardo Sturani$^{b,c}$
\par}
{\small\it\centering\ignorespaces

(a) INFN, sezione di Pisa, Pisa, Italy\\
(b) Dipartimento di Fisica dell'Universit\`{a},
piazza Torricelli 2, I-56100 Pisa, Italy\\
(c) Scuola Normale Superiore, 
piazza dei Cavalieri 7,I-56125 Pisa, Italy.\par}

\par
\bgroup
\leftskip=0.10753\textwidth \rightskip\leftskip
\dimen0=-\prevdepth \advance\dimen0 by17.5pt \nointerlineskip
\small\vrule width 0pt height\dimen0 \relax

We examine the effect of spatial curvature in the pre-big-bang
inflationary model suggested by string theory. 
We study  $O(\alpha ')$ corrections and we show that,
independently of the initial curvature, they lead to a  phase
of exponential inflation. The amount of inflation in this phase
is long enough to solve the
horizon and flatness problems if  the evolution starts 
deeply into the weak coupling regime. There is
 a region of the parameter space of the model where such a
long inflationary phase at the string scale is consistent with 
COBE anisotropies, millisecond pulsar timing and nucleosynthesis
constraints. We discuss implications for  the spectrum of relic
gravitational waves at the frequencies of LIGO and Virgo.

\par\egroup

\thispagestyle{plain}
\endgroup
\vspace{5mm}

String cosmology~\cite{Ven,GV,review} 
provides a possible implementation of
the inflationary paradigm with two major advantages. First, it
addresses the problem of the initial singularity and, second, the
`inflaton'  is identified with a  field, the dilaton,
whose dynamic is not prescribed {\em ad hoc}, but rather follows from a
fundamental theory. 

A principal motivation of any inflationary model is to get rid of
the fine tuning of the initial conditions of standard cosmology. 
In a recent paper~\cite{TW} the issue of the dependence
on the initial conditions in string cosmology
has been reanalyzed,  with the conclusion that 
a fine-tuning  is still required to
obtain enough inflation to solve the horizon/flatness problems. The
analysis of ref.~\cite{TW} focused on the super-inflationary
phase of the model (see below).
In this Letter we consider the effect of
spatial curvature on the `string' phase of the model, and the contribution
of this phase to the solution of the fine tuning problem.

The model is based on the low-energy effective action of string theory
and depends on the metric $g_{\mu\nu}$ and on the dilaton field
$\phi$. The effective action is given by an expansion in powers of the
string constant $\alpha '$; including the first order $\alpha '$
correction  it can be written (in the so-called `string frame') as
\be\label{act}
S =-\frac{1}{2\ls^2}
\int d^4x\sqrt{-g}e^{-\phi}\left[ R+(\nabla\phi )^2
-\frac{\alpha '}{4}\left( R_{\rm GB}^2-(\nabla\phi )^4\right)
\right]\, ,
\ee
where $\ls$ is the string length, $\alpha '\sim\ls^2$ is the string
constant  and $ R_{\rm GB}^2$ is the Gauss-Bonnet term.
Our sign conventions are $\eta_{\mu\nu}=(+,-,-,-)$ and
$R^{\mu}_{\hspace*{1mm}\nu\rho\sigma}
=\partial_{\rho}\Gamma^{\mu}_{\nu\sigma}+\ldots$.
Further terms in the action include an antisymmetric tensor field,
higher order corrections in $\alpha '$, corrections
$O(e^{\phi})$, a non-perturbative dilaton potential
that, as $\phi\ra -\infty$, vanishes as a double exponential, 
matter and gauge fields, and moduli fields. 

Out of all these terms, 
$O(\alpha ')$ and $O(e^{\phi})$ corrections are particularly important
and play a crucial role in
defining a consistent model. This is most easily understood
considering the special case  of a spatially flat
Friedmann-Robertson-Walker (FRW) metric:  in this case
the evolution starts at $t\ra -\infty$
at low curvature and $\phi\ra -\infty$,
where both $O(\alpha ')$ and $O(e^{\phi})$
corrections are neglegible. In this regime
we can neglect the $\sim\alpha '$ term in eq.~(\ref{act}) and
 the evolution of the FRW scale factor $a(t)$ is
$a(t)\sim 1/(-t)^{\gamma}$ with $\gamma >0$. This corresponds
to super-inflation, and the solution runs into a singularity as 
$t\ra 0^-$. However, at a value of time 
$t=t_s<0$, the curvature becomes of
order $1/\ls^2$ and  $\alpha '$ corrections
become crucial. Let us assume that at this 
 point $e^{\phi}$ 
is still small,  $g_s^2=e^{\phi_s}\ll 1$. Then $O(e^{\phi})$
corrections can still be neglected.
Once we include $\alpha '$ corrections the solution, rather than
running into the singularity, approaches asymptotically a 
stage of exponential inflation 
with a linearly growing dilaton,
$a(t)\sim\exp\{ H_st\},\phi (t)=\phi_s +c(t-t_s)$,
with $H_s,c$  constants of order $1/\ls$.
In terms of conformal time $\eta$, this solution reads
\be\label{sol}
a(\eta )=-\frac{1}{H_s\eta}\, ,\hspace{5mm}
\phi(\eta )=\phi_s-2\beta \ln\frac{\eta}{\eta_s}\, ,
\ee
where $\eta_s=\eta (t_s)$,  $\eta_s <\eta <0$ and  $2\beta =c/H_s.$
This solution exists
at all orders in $\alpha '$ if a
set of two algebraic equations that determine the values of $H_s,c$,
and that involves all orders in $\alpha '$, has real
solutions~\cite{GMV}. We will assume in the following that this is
indeed the case, and
we will refer to this phase as DeSitter phase,
or string phase. Together with the previous super-inflationary
phase (or `dilaton dominated' phase) it defines the so-called
`pre-big-bang cosmology'.\footnote{Recently one of us has discussed a
different mechanism for the  regularization of the singularity, based
on the production of massive string modes. In this case the
Hubble parameter  in the string phase is not constant, but rather
bounces back after reaching a maximum value. The detailed evolution in
this case is sensitive to both $\alpha '$ and $O(e^{\phi})$
corrections, and we hope to report on it in future work. The scenario
discussed in the present paper is appropriate if the value of $H_s$,
determined by perturbative $\alpha '$ corrections, is smaller than the
typical value of $H$ where the production of massive modes sets 
in~\cite{MM}.}
The numerical
value of $\beta$ is very difficult to compute since it is sensitive
to all higher order corrections in $\alpha '$. It also depends on the
dimensionality of space-time, i.e., on whether the compactification of
extra dimensions takes place before or after the string phase. We must
therefore consider it as a free parameter of the model, although it is
in principle fixed by the theory; its numerical value is
relevant  in the following, and it is
also important  when considering the phenomenological consequences 
of the model. In fact
pre-big-bang inflation predicts a relic
graviton spectrum which is particularly interesting, for the detection
in planned gravitational wave experiments, 
 if $|2\beta -3|$ is very close to 3, i.e., for
$\beta \simeq 0$ or $\beta\simeq 3$~\cite{BGGV,BMU}. This gravitational
wave spectrum also puts important  constraints on the model, and we 
will discuss them in some detail below.

Another very important parameter of the model is the value $\phi
(\eta_s)=\phi_s$ of the dilaton at
the transition between the super-inflationary and the DeSitter phases.
Note that $g^2=e^{\phi}$ plays the role  of the gauge coupling
`constant' (which actually is only constant in the present era, when
the dilaton is frozen at a minimum of the nonperturbative potential).
As we have already remarked, the solution~(\ref{sol}) is valid
provided $g_s^2\ll 1$.
The string phase 
is expected to end and to match with the standard radiation dominated era
(`graceful exit') when $e^{\phi}\sim O(1)$, although the detailed
mechanism is  not yet well understood~\cite{gra1,gra2}. 

The condition $g_s^2\ll 1$ ensures the existence of a long string
phase, whose consequences we will explore below. We also note that
a long intermediate string phase might also play an
essential role in the generation of density perturbations necessary
for the formation of large scale structures~\cite{MS}, and
  that the assumption that the Universe starts deeply within
the weak coupling region is also necessary in order to relax the
hypotesis of homogeneity  of the initial conditions~\cite{Ven2}.

Let us now turn to the generic case of a spatially closed ($k=1$) or a
spatially open ($k=-1$) Universe. The equations of motion for
homogeneous fields derived from the action~(\ref{act}) 
with a variation with respect to the scale factor $a(t)$ and with
respect to the dilaton field read
\bees\label{dyn1}
2\frac{\ddot a}{a}+\frac{\dot a^2}{a^2}+\frac{k}{a^2}+\frac{\dot \phi^2}
{2}-\ddot\phi-2\frac{\dot a}{a}\dot\phi-\alpha'\left[2\frac{\dot a}{a}
\frac{\ddot
a}{a}\dot\phi+\frac{1}{8}\dot\phi^4+\left(\ddot\phi-\dot\phi^2
\right)\left(\frac{\dot a^2}{a^2}+\frac{k}{a^2}\right)\right]=0\\
\label{dyn2}
-6\left(\frac{\ddot a}{a}+\frac{\dot a^2}{a^2}+\frac{k}{a^2}\right)-
\dot\phi^2+2\ddot\phi+6\frac{\dot a}{a}\dot\phi+3\alpha'\left[-2\frac
{\ddot a}{a}\left(\frac{\dot a^2}{a^2}+\frac{k}{a^2}\right)-\frac{1}{4}
\dot\phi^4+\dot\phi^3\frac{\dot a}{a}+\dot\phi^2\ddot\phi\right]=0
\ees
The variation with respect to the lapse function gives a constraint on
the initial values,
\be\label{constraint}
\dot\phi^2+6\left(\frac{\dot a^2}{a^2}+\frac{k}{a^2}\right)-6\frac{\dot a}
{a}\dot\phi-3\alpha'\left[2\dot\phi\frac{\dot a}{a}\left(\frac{\dot a^2}
{a^2}+\frac{k}{a^2}\right)-\frac{1}{4}\dot\phi^4\right]=0
\ee
that is preserved by the dynamical equations~(\ref{dyn1},\ref{dyn2}).
Without the $\alpha '$ corrections, these equations have been discussed
in~\cite{CLW,altri,TW}.  For a spatially closed
Universe, $k=1$, the solution cannot be extrapolated back in time to
$t\ra -\infty$, because it hits a singularity (close to the
singularity, of course, $\alpha '$ corrections cannot be
neglected). The solution with $k=-1$, instead, can be extrapolated to
$t\ra -\infty$ but at large negative values of time the scale factor
is very large and increases as we go backward in time,
and this might be an  implausible initial condition. 

\begin{figure}[t]
\centering
\includegraphics[width=0.4\linewidth,angle=270]{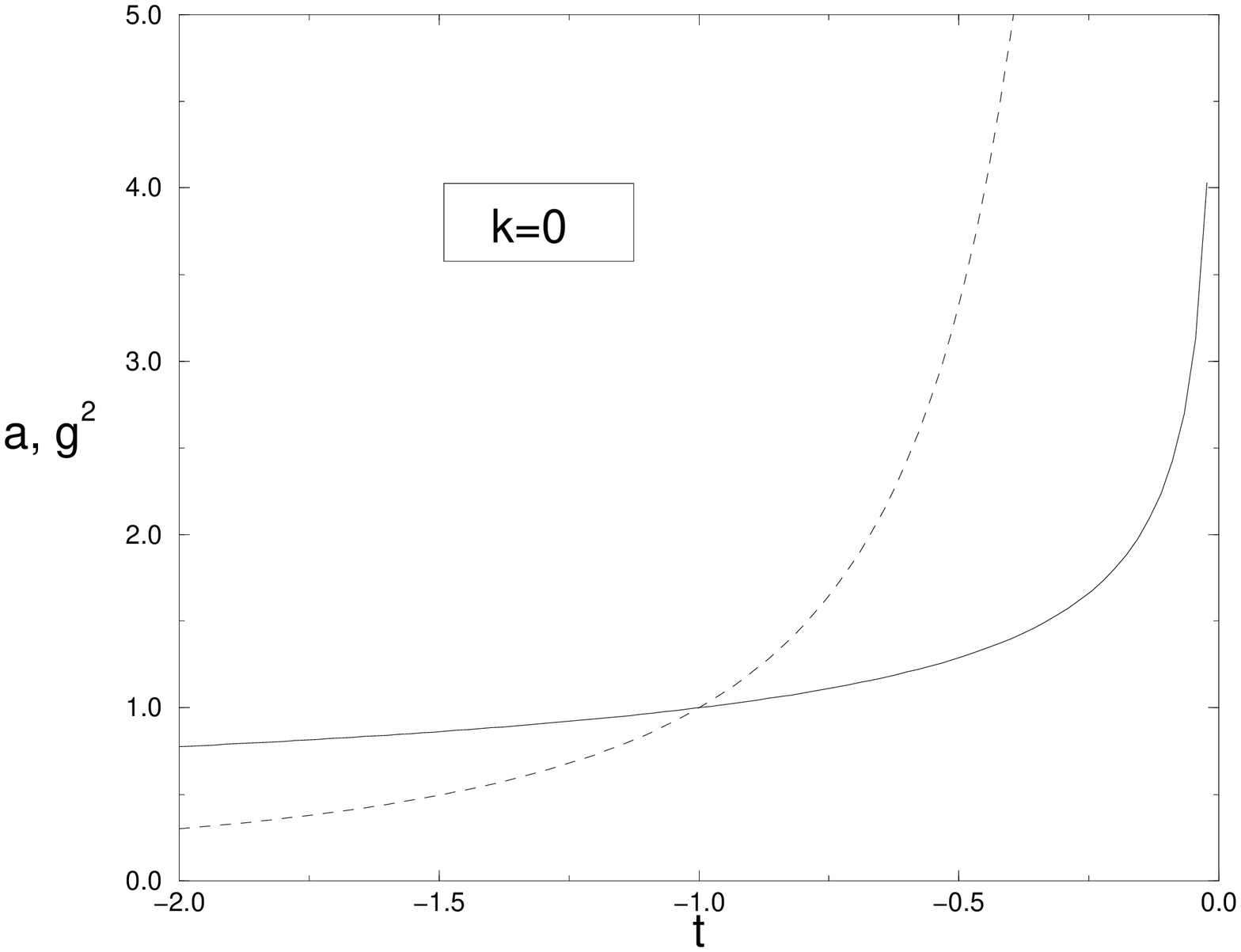}
\includegraphics[width=0.4\linewidth,angle=270]{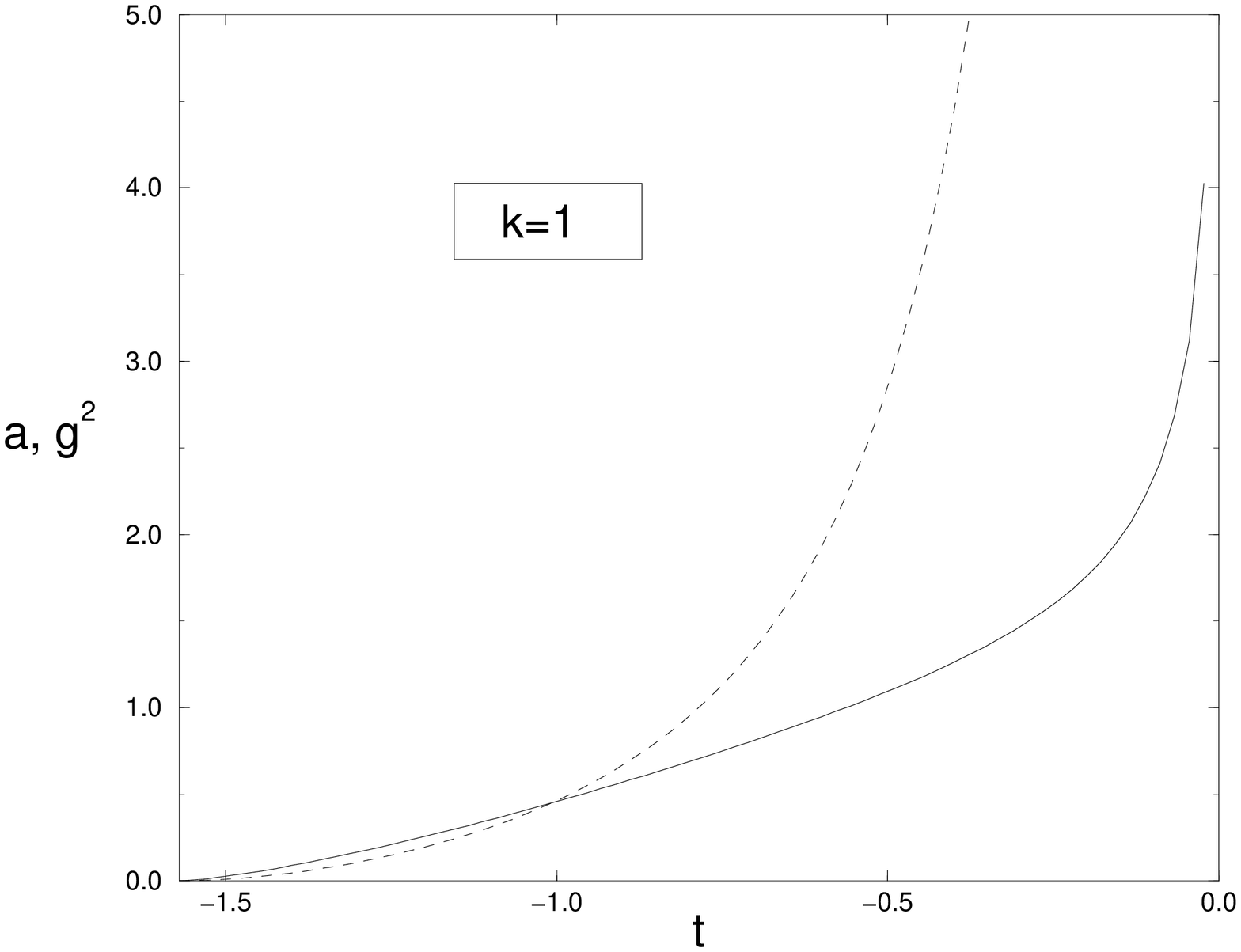}
\mbox{(a)}\mbox{\hspace{9cm}}\mbox{(b)}
\includegraphics[width=0.4\linewidth,angle=270]{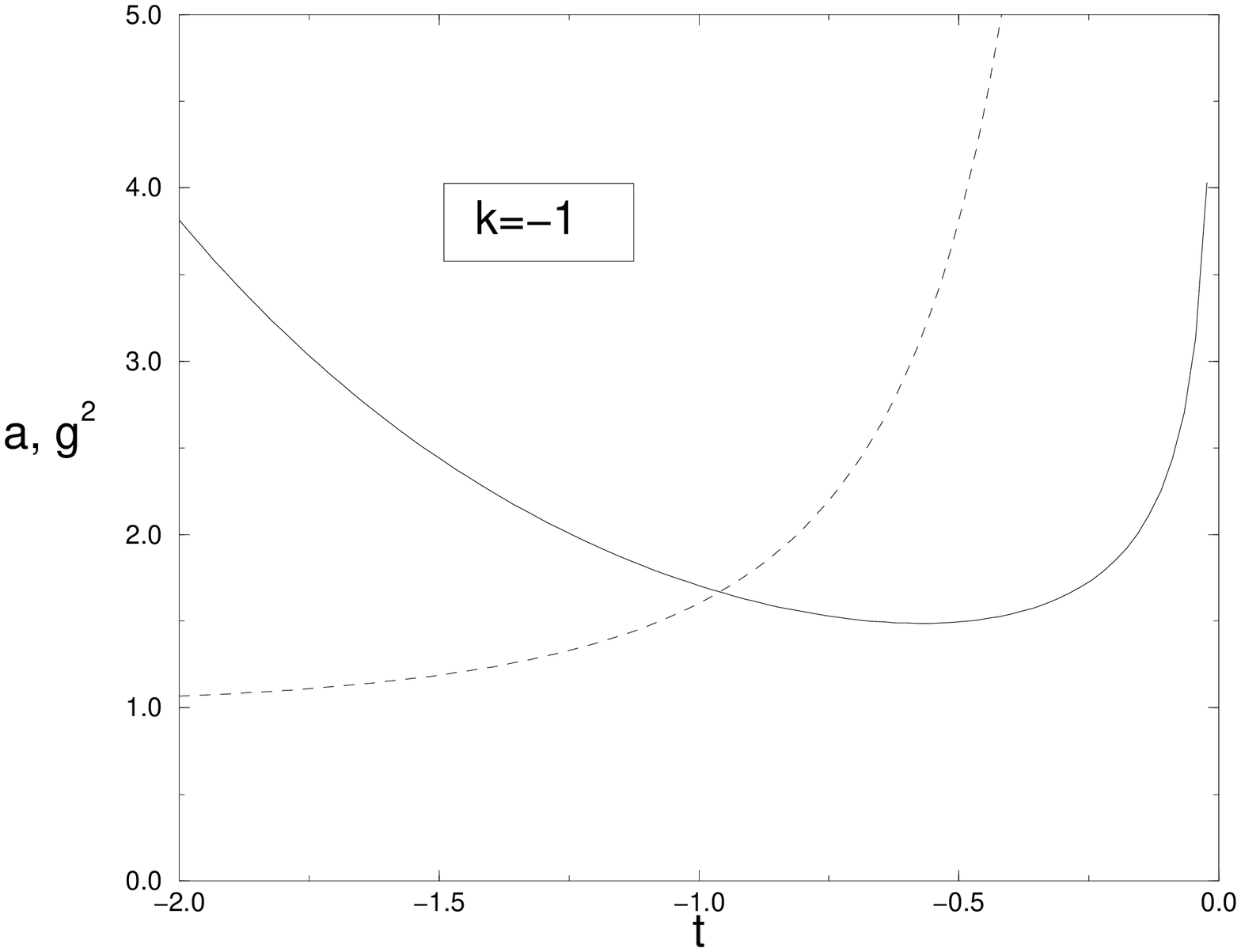}

\mbox{(c)}
\caption{Evolution of the cosmic scale factor $a$ (solid line) and
of the gauge coupling $g^2\equiv e^\phi$ (dashed line) from eqs.
 (\ref{dyn1},
\ref{dyn2}) with $\alpha'\equiv 0$ for the three possible cases of 
spatial curvature. The scale of the vertical axes
are arbitrary, depending on the values of
$a_0$ and $\phi_0$.}
\end{figure}

Following the strategy of ref.~\cite{TW}, we will therefore
assign the initial conditions at some finite (negative) value of
time $t_0$, not very close to the would-be singularities.
Physically, one expects that because of quantum fluctuations, at
some value of $t=t_0$ a sufficiently smooth patch emerges, which then
starts inflating~\cite{Ven2}.

The initial conditions for eqs.~(\ref{dyn1},\ref{dyn2}) are specified by
$a_0,\dot{a}_0,\phi_0,\dot{\phi}_0$. However, $\phi_0$ never
enters the equations, as can be seen from the fact that, in the
absence of external matter fields in the action, a 
constant shift in $\phi$
gives simply an overall multiplicative factor in the
action~(\ref{act}) and therefore does not affect the classical
equations of motion; thus we have a family of solutions differing only
by a shift in $\phi$.
Furthermore, $a_0,\dot{a}_0,\dot{\phi}_0$ are not independent, because
of the constraint, eq.~(\ref{constraint}). We will take
$a_0$ and $H_0=\dot{a}_0/a_0$ as independent variables in the following.
The physical curvature radius of a FRW
Universe is~\cite{KT} $R_{\rm curv}=a(t)/|k|^{1/2}$ and the spatial
curvature is $^3{\cal R}=6/R_{\rm curv}^2$. 
Since we have performed the usual rescaling so that
for a spatially curved space $|k|=1$, then $a_0$ is the
initial curvature radius. The spatially flat limit is recovered as
$a_0\ra\infty$. 

We have  studied eqs.~(\ref{dyn1},\ref{dyn2})  for different values of
$k,a_0,H_0$. Figs.~(1a,1b,1c) show the result of the integration
for $a(\eta )$ and for the effective gauge coupling 
$g^2(\eta )=\exp\{\phi (\eta )\}$, versus conformal time $\eta$,
without  $\alpha '$ corrections, in the three cases  
$k=0,k=\pm 1$~\cite{CLW}. At a finite value of $\eta$ the
solution hits a singularity. Close to the singularity, the solutions
for the three cases approach each other, as can be seen
analytically. 

\begin{figure}
\centering
\includegraphics[width=0.4\linewidth,angle=270]{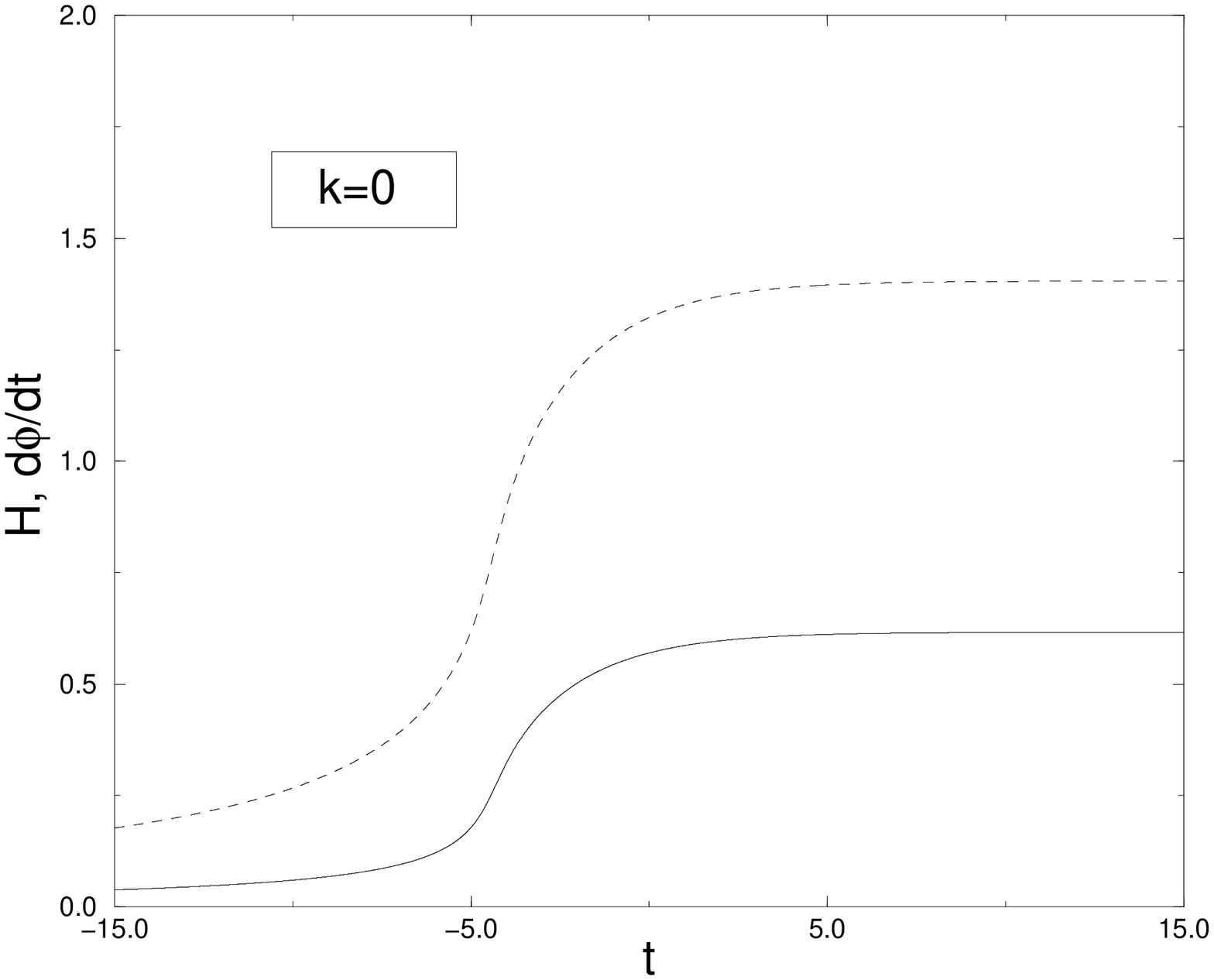}
\includegraphics[width=0.4\linewidth,angle=270]{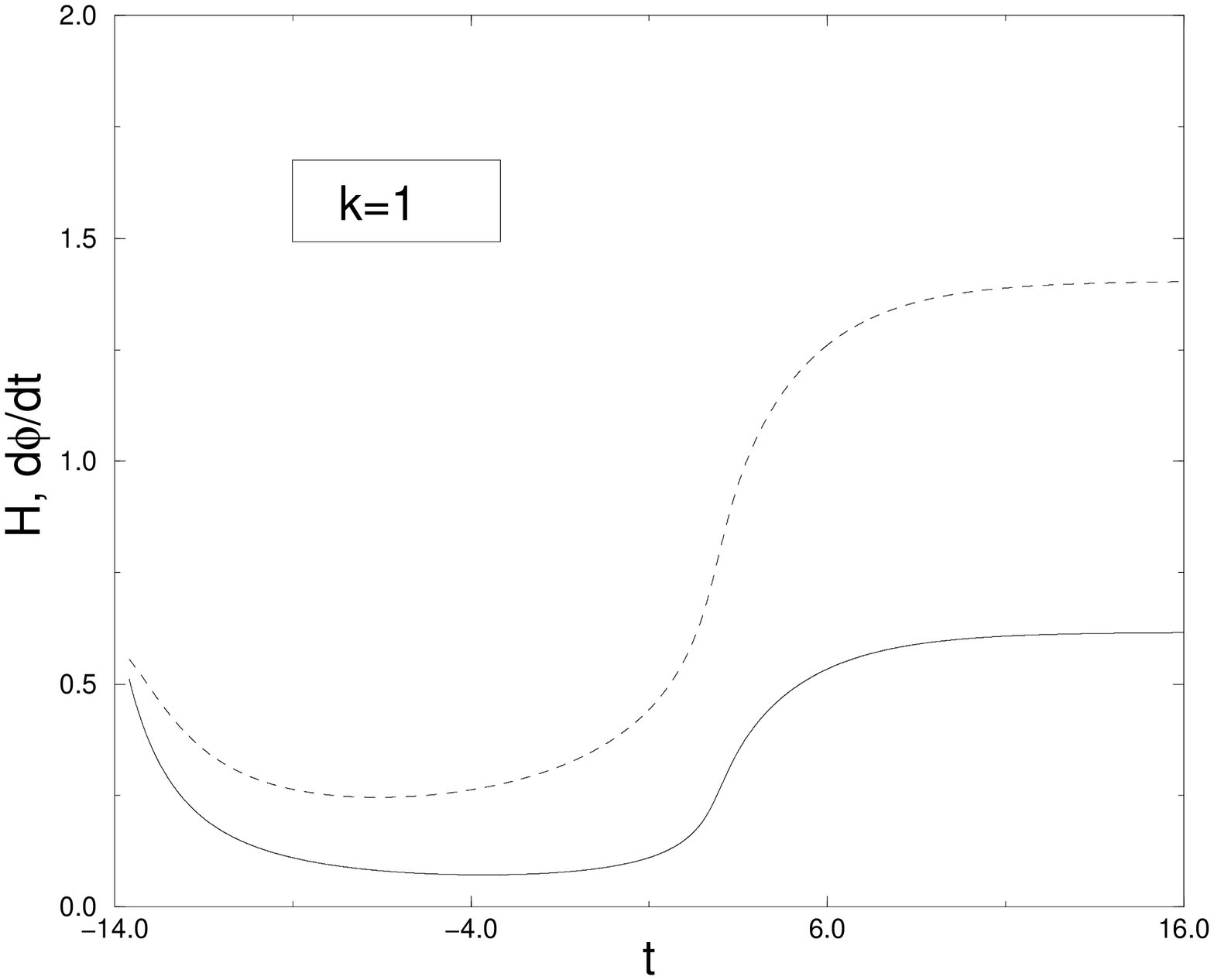}
\mbox{(a)}\mbox{\hspace{9cm}}\mbox{(b)}
\includegraphics[width=0.4\linewidth,angle=270]{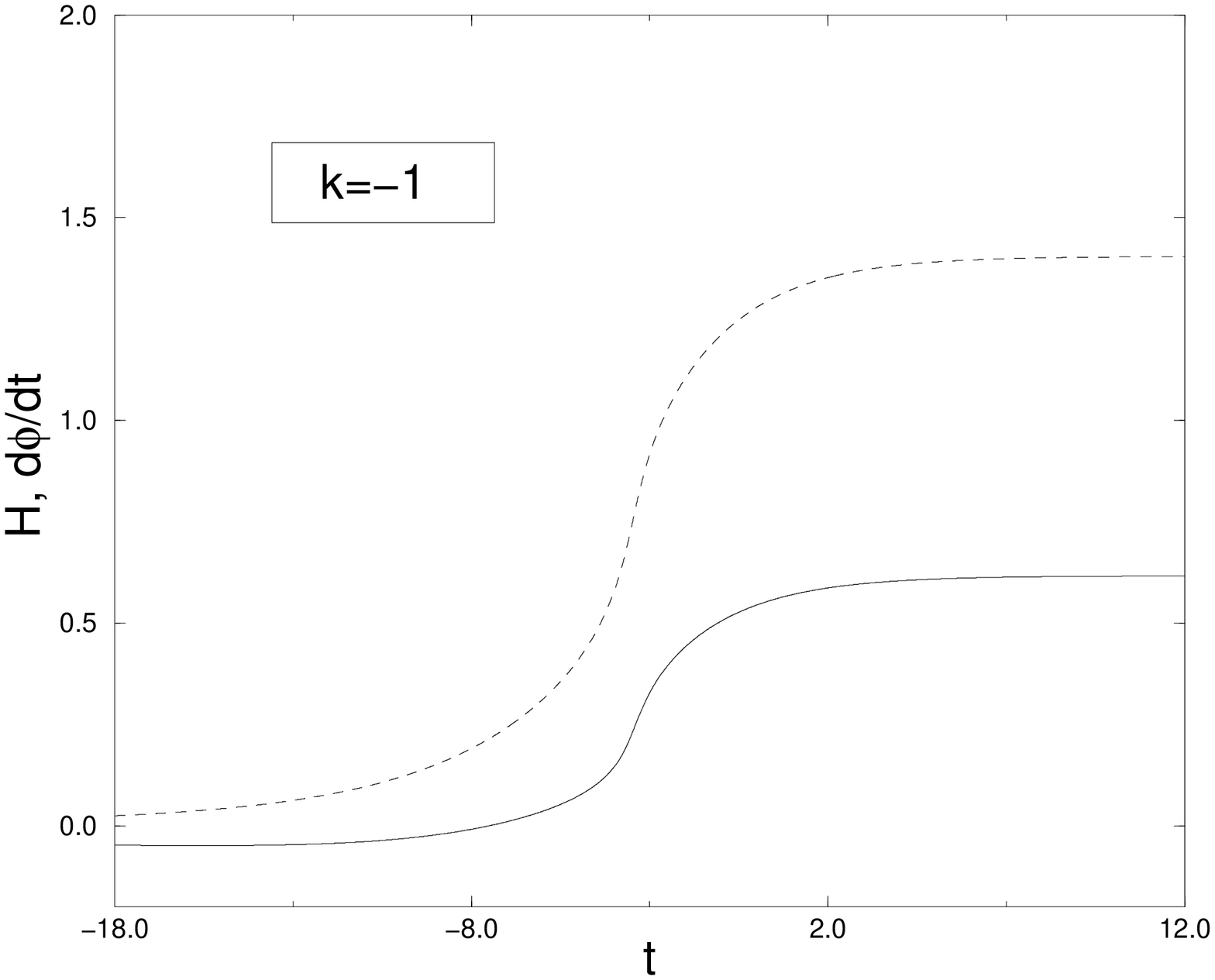}

\mbox{(c)}
\caption{Evolution of the Hubble parameter $H$ (solid line) and of the
derivative of the dilaton with respect to
 cosmic time $\dot\phi$ (dashed line) 
from eqs. (\ref{dyn1},\ref{dyn2}), including
 $\alpha '$ corrections, for the three possible cases of 
spatial curvature. The  vertical axes are in units of 
$1/\sqrt{\alpha '}$. The
initial conditions $H_0$ and $\dot\phi_0$ are chosen so that they lie on the
pre-big-bang solutions with $\alpha '=0$ exhibited in figs. (1a,1b,1c).}
\end{figure}

Figs. (2a,2b,2c) show the result of the numerical integration
including $O(\alpha ')$ corrections,  for $k=0,k=\pm 1$. We plot
$H$ and $\dot{\phi}$ vs. $\eta$,
where $H$ is the Hubble parameter, $H=\dot{a}/a$.  We see that
in all three cases the $\alpha '$ corrections regularize the lowest
order solution,
 and in all three cases we have a phase of exponential
inflation with a linearly growing dilaton, $H\simeq
H_s,\dot{\phi}\simeq c$.
Furthermore the asymptotic values  $H_s,c$ are the
same in the three cases. This can also be easily seen analytically,
since in eqs.~(\ref{dyn1},\ref{dyn2}) $k$ always appears in the combination
$(\dot{a}/a)^2+k/a^2$. During superinflation, the term $(\dot{a}/a)^2$
becomes much bigger than $1/a^2$ as $t$ approaches zero, and during
exponential inflation $(\dot{a}/a)^2$ continues to increase compared to 
$1/a^2$, and then the term $k/a^2$ becomes quickly irrelevant.

The total amount of inflation between two generic values of
time $t_i$ and $t_f$ can be
conveniently measured by the factor
\be\label{Z}
Z=\frac{H(t_f)a(t_f)}{H(t_i)a(t_i)}\, ,
\ee
The total amount of inflation
during the DeSitter phase, for a spatially flat space,
is  easily found from eq.~(\ref{sol}), 
fixing the end of the inflation from the
condition $\phi (t_f)=0$:
\be\label{ZDS}
Z_{\rm DS}=\exp\left\{\frac{|\phi_s|}{2\beta}\right\}\, ,
\ee
and it is very large if at the beginning of the string phase we are
in the weak coupling regime, $|\phi_s|\gg 1$, or if $\beta \ll 1$.

Since the scale factor and dilaton field during the string phase are
still given by eq.~(\ref{sol})  also for $k=\pm 1$, 
at least within our numerical precision, the
amount of inflation in the string phase  is  still given by
eq.~(\ref{ZDS});
furthermore, even the constant $2\beta =c/H_s$ that 
appears in eq.~(\ref{ZDS}) is  the  same for $k=0$ and for $k=\pm 1$.

\begin{figure}
\centering
\includegraphics[width=0.55\linewidth,angle=270]{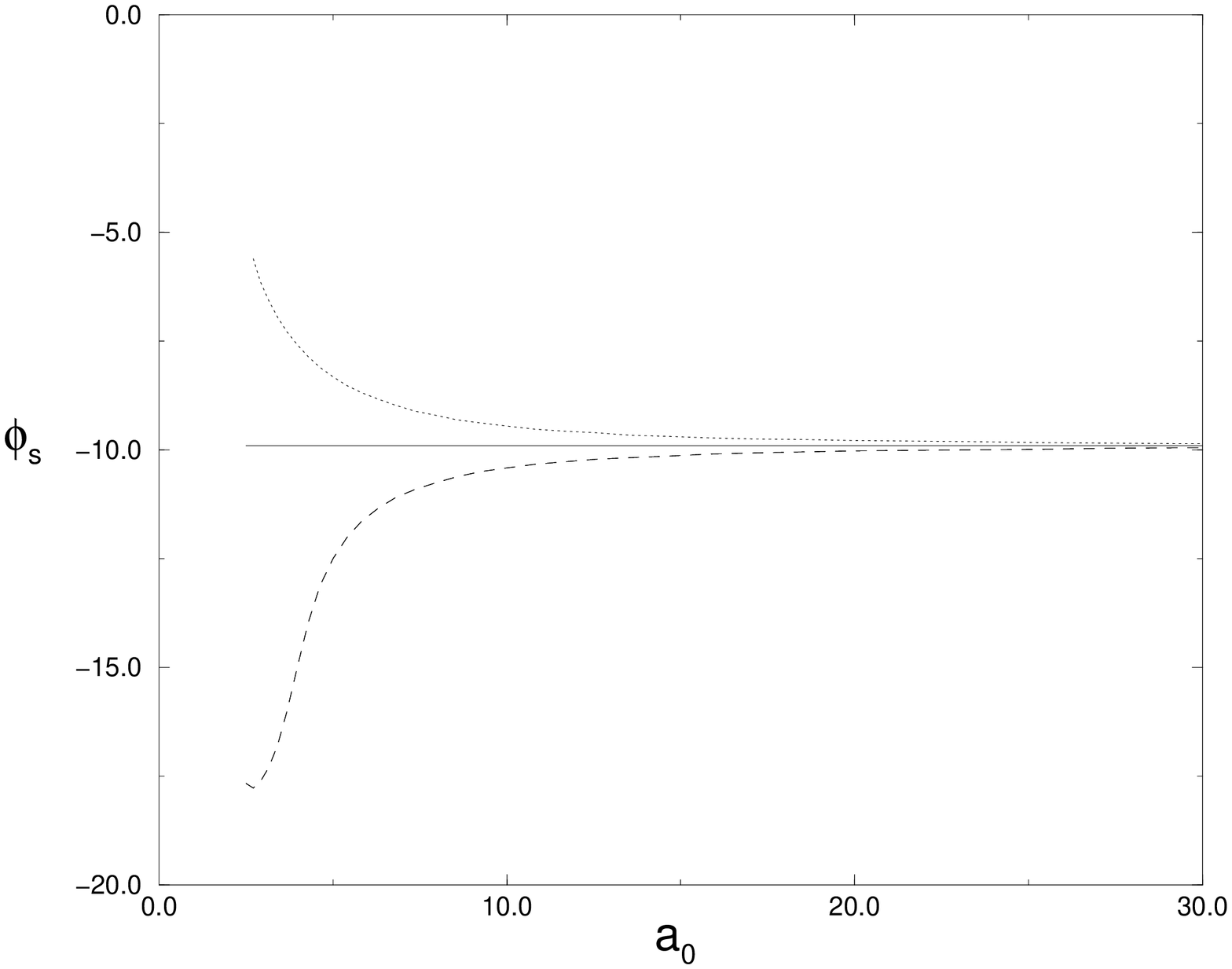}
\caption{$\phi_s$ (defined as the value $\phi$
when both $H$ 
and $\dot\phi$ are constant within an accuracy of one per cent) 
as a function of
$a_0$ (with $H_0=0.5$ and $\phi_0=-20$) 
for $k=0$ (solid line), 
$k=1$ (dashed line) and $k=-1$ (dotted line).}
\end{figure}

To understand the dependence of $Z_{\rm DS}$ on the initial conditions
we therefore
study $\phi_s(a_0,H_0,\phi_0,k)$. Since the equations of motion 
depend on $\dot{\phi}$ and not on $\phi$, $\phi_s$ has the general 
form
\be
\phi_s(a_0,H_0,\phi_0,k)=\phi_0+f(a_0,H_0,k)\, .
\ee
The function $f(a_0,H_0,k)$  is shown in fig.~3 for the three values of
$k$ vs. $a_0$ for fixed
$H_0$ (actually, for $k=0$, $f$ does not depend on $a_0$, so we can 
introduce $\tilde{f}(H_0)\equiv f(a_0,H_0,0)$).
Each
point in the graph is the result of a numerical integration with 
the given values of $a_0,H_0$ as initial condition; $\phi_s$ is
defined, operatively, as the value of $\phi$ when both $H$ 
and $\phi$ are constant within an accuracy of one per cent.
For large values of $a_0$, $f(a_0,H_0,k=\pm 1)$ 
approaches smoothly its flat
space limit,  $f(\infty ,H_0,k=\pm 1)=\tilde{f}(H_0)$. 
In fig.~3, where we have fixed $H_0=0.5/\sqrt{\alpha '}$, the value of
$|\phi_s|$  for $k=1$ is larger 
than the value for $k=-1$, starting with the same $\phi_0$.  For
smaller values of $H_0$ we have found that
the situation is reversed and $|\phi_s|$ is larger for $k=-1$.

We now ask when the amount of inflation during the string phase 
alone, given by
eq.~(\ref{ZDS}), is sufficient to solve the cosmological problems. 
The solution of the horizon problem requires~\cite{KT} $\ln Z>O(60)$,
and therefore in our case we must have
\be\label{cond}
|\phi_s|\gsim 120\beta\, .
\ee

In general, we can expect that a long inflationary phase at the string
scale will produce a large density of stochastic gravitational waves,
and we should ask whether the condition~(\ref{cond}) is consistent
with the experimental bounds on the gravitational wave spectrum. The
spectrum is conveniently characterized  by the quantity
\be
\Omega_{\rm GW}(f)=\frac{1}{\rho_c}\,
\frac{d\rho_{\rm GW}}{d\ln f}\, ,
\ee
where $f=\omega/(2\pi )$ is the frequency, $\rho_{\rm GW}$ is the
energy density in gravitational waves and $\rho_c$ is the critical
density of the Universe. The quantity to be compared with the
experimental results is actually $h_0^2\Omega_{\rm GW}$, where $h_0$
is the uncertainty on the Hubble constant, $H_0=h_0 100 {\rm
km/(sec-Mpc)}$, since $h_0^2\Omega_{\rm GW}$
 is independent of the uncertainty in the
quantity $\rho_c$ that we use to normalize $\rho_{\rm GW}$.
The main observational bounds for the spectrum 
are~\cite{BGGV,KT,Al}:
\begin{itemize}
\item the COBE bound
\be
h_0^2\Omega_{\rm GW}(f)<7\times 10^{-11}\left(\frac{H_0}{f}\right)^2
\qquad\mbox{for}\quad H_0<f<30 H_0,
\ee
\item the millisecond pulsar timing constraint
\be
h_0^2\Omega_{\rm GW}(f=10^{-8}Hz)<10^{-8},
\ee
\item the compatibility with the standard nucleosynthesis 
scenario
\be\label{ns}
\int \Omega_{\rm GW}(f)d(\ln f)\leq 6.3\times 10^{-6}
\ee
\end{itemize}
(in the last equation we have used the value $N_{\nu}<3.9$ for the
equivalent number of neutrino species, see~\cite{Cop}).
The spectrum of gravitational waves predicted by string 
cosmology~\cite{BGGV,BMU}
is characterized by two parameters $f_1,f_s (f_s<f_1)$ with dimension of
frequency, and by the dimensionless constant $\beta$; 
the spectrum is an increasing function of $f$ up to 
the cutoff frequency $f_1$ which is of the order of 10 GHz.
The parameter $f_s$ is related to $\phi_s,\beta$ by
$Z_{\rm DS}=f_1/f_s$, where $Z_{\rm DS}=\exp\{ |\phi_s|/(2\beta )\}$,
see eq.~(\ref{ZDS}).

In the range $f_s<f<f_1$ the spectrum is
approximately\footnote{We neglect here the finer details of the
spectrum, which are discussed in ref.~\cite{BMU}.} given by
\be
h_0^2\Omega_{\rm GW}(f)\simeq 3\times 10^{-7}\left(\frac{f}{f_1}
\right)^{3-|3-2\beta |}\, .
\ee
while for $f<f_s$ it varies as $f^3\ln^2f$.
This form of the spectrum has been computed  for
$k=0$. We have repeated the computation for 
$k=\pm 1$~\cite{RS} and we find that, with very high accuracy, 
it is  unchanged.  For wavelengths much
smaller than the present Hubble radius of the Universe, i.e. for $f\gg
H_0$, this is obvious, since the gravitational wave does not feel
the structure of the Universe on a length scale much bigger than its
wavelength. For sufficiently large wavelengths one should expect a
modification of the spectrum, expecially for
a closed Universe, because in this case
the modes of the gravitational field are  discrete~\cite{BD};
however the spacing is $\Delta f_{\rm com}=1/(2\pi )$, 
where $f_{\rm com}$ is the comoving frequency.
This corresponds to a spacing in the physical frequencies today
$\Delta f=1/(2\pi a(t_{\rm pres}))$, where $a(t_{\rm pres})$ is the
present value of the scale factor, in units $k=-1$.  In general, also
for an open Universe, this is the frequency scale where the spectrum is
modified compared to the case of flat space. However, because of the
inflationary evolution, $1/a(t_{\rm pres})$ is much smaller
 than $H_0$, and
therefore at COBE frequencies, $f\sim H_0$,  the spectrum
is still indistinguishable from the flat space case,
$k=0$.

Both the frequency $f_1$ and the peak value $h_0^2\Omega_{\rm GW}(f_1)$
are fixed, either using the so-called one-graviton level~\cite{BGGV}, 
or using the
expected value for the string mass scale and therefore for
$H_s$~\cite{BGGV,BMU},  in the range
$f_1=$ 6--20 GHz, $h_0^2\Omega_{\rm GW}(f_1)=$(0.1--8)$\times 10^{-7}$. 
In the following, for definiteness, we use
 $f_1=10\, {\rm GHz}$ and $h_0^2\Omega_{\rm GW}
(f_1)=3\cdot 10^{-7}$, as in \cite{BMU}.  Our results can be easily
rescaled using  different values for these quantities.

Since $f_1$ is fixed,
the spectrum depends only on two free parameters $\beta, f_s$,
or equivalently $\beta ,\ln Z_{\rm DS}$.
We now study the range
of values of these parameter allowed by 
the three observational constraint discussed above.

We insert the explicit form of the spectrum in order to compute the
integral in the bound~(\ref{ns}); combining it
with the COBE and pulsar bounds we obtain the results presented in
fig.~4. The shaded area is the region of parameter space forbidden by
these observational constraints. 
We observe that if we require $\ln Z_{\rm DS} >60$ 
we must have $\beta >0.12$
in order to evade the COBE bound. This is due to the fact that with 
$\ln Z_{\rm DS} >60$ the frequency $f_s=f_1/Z_{\rm DS}$ becomes
smaller than the maximum frequency explored by COBE
$f\simeq 10^{-16}$ Hz, and  we cannot
take advantage of the $\sim f^3$ behavior of $\Omega_{\rm GW}$ for
$f<f_s$ in order to
lower the value at COBE frequencies in comparison to the value at
$f=f_1$. Rather, from $10^{-16}$ Hz up to $f=f_1\sim 10$ GHz the
spectrum varies as $f^{3-|3-2\beta |}$ (i.e. as
$f^{2\beta}$ for $\beta \leq 3/2$), and, because of this, $\beta$ cannot be
too close to zero. 

We can also consider a situation in which the amount of inflation 
$\ln Z >60$ is given partly by the DeSitter phase and partly by the
super-inflationary phase. This reduces 
the requirement on $Z_{\rm DS}$ alone and therefore on $\beta$.
A value of $\beta$ as close as possible to zero is the most
favorable situation for the observation of the gravitational wave
spectrum at LIGO/Virgo frequencies, $f=$ 6Hz--1kHz. Therefore, if we
ask that the required amount of inflation is provided uniquely by the
string phase, the maximum value of the spectrum at, say, 1kHz, is
lowered. 
We present in the figure  the lines in the
parameter space that correspond to a value of 
$h_0^2\Omega_{\rm GW} (1\,{\rm kHz})$
equal to $10^{-9},10^{-8}$ and $10^{-7}$, respectively.

In the range $40<\ln Z<56$ the stronger limit on $\beta$ is given by
 the pulsar-timing constraint, and is $\beta> 0.04$.
Finally for $21<\ln Z<40$ the primordial nucleosynthesis constraint 
is the strongest one and we find the final smooth branch of the curve.
For $\ln Z<21$ we have no more restrictions on the value of $\beta$.

\begin{figure}
\centering
\includegraphics[width=0.6\linewidth,angle=270]{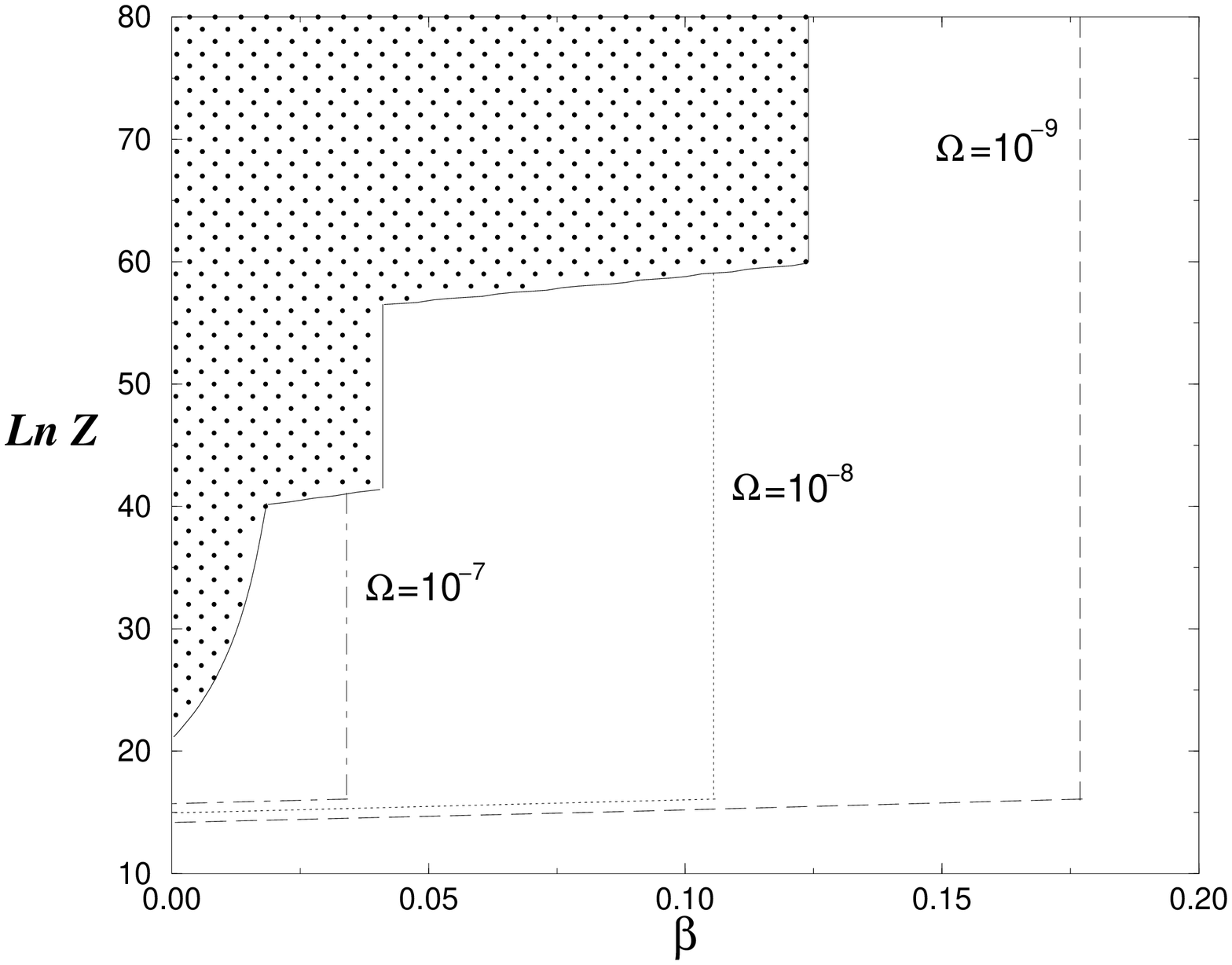}
\caption{The forbidden  region in the parameter space is the shaded
area.
Along the dot-dashed line $\Omega=h_0^2\Omega_{\rm GW}(1\,{\rm kHz})=
10^{-7}$, along the dotted one $\Omega =10^{-8}$ and along the dashed line
$\Omega =10^{-9}$.}
\end{figure}

To understand the issue of fine tuning, it is also useful to
discuss our results in terms of the original parameters of the model
$\phi_s$ and $\beta$, rather than $\ln Z_{\rm DS}$ 
and $\beta$. A large value of $Z_{\rm DS}=-\phi_s/(2\beta )$ can 
be obtained as a combination of two limiting cases:
(i)  if $\beta$ is very close to
zero, so that for any reasonable value of the initial curvature
eq.~(\ref{cond}) is satisfied, even without requiring an especially
large value of $|\phi_0|$ and hence of $|\phi_s|$; or
(ii) if $\phi_0$ is very large and negative; 
in this case  the dependence of $\phi_s$ on  
$a_0$ is neglegible for natural values of $a_0$, and 
again the amount of inflation is sufficiently large.

Concerning condition (i), we see from fig.~4 that we cannot choose
$\beta$ arbitrarily small because of the various observational
constraints; still, we can reach moderately small values $\beta\simeq
0.15$. 
Condition (i) is  analogous to the slow roll conditions in
standard implementations
of the inflationary scenario. It is a requirement on the dynamics of
the theory, not on the initial conditions, and it ensures 
that $\dot{\phi}$ is sufficiently small
so that the mechanism that terminates inflation, and that 
presumably takes place when $e^{\phi}=O(1)$, is sufficiently delayed.
A solution of the fine tuning problem based on the option (i) is
therefore very similar, conceptually, to what is done in
other  inflationary models. 

String cosmology, however,  also has the option (ii). In this case the
inflationary phase is long not because the field $\phi$ obeys a
slow-roll condition, but rather because its initial value $\phi_0$ is
such that $e^{\phi_0}\ll 1$ is very far from the point where inflation
terminates, $e^{\phi}\sim 1$. We can compare this with what happens in
chaotic inflation~\cite{Lin}. In this case the `natural'
initial value of the
inflaton field $\varphi_0$ is fixed by the condition $V(\varphi_0)
\sim M_{\rm Planck}^4$, where $V$ is the potential that triggers
inflation, and this fixes the dimensionful field $\varphi_0$ in terms
of the Planck mass and of the dimensionless parameters of the potential. 

In our case, instead, $\phi$ is a dimensionless field and 
 $g_0^2=e^{\phi_0}$ is the initial value of the gauge coupling.
The initial condition $g_0^2\ll 1$ means that the evolution starts
deeply into the perturbative regime. As such,
we do not regard $g_0^2\ll 1$
 as a fine-tuned initial condition; rather, it is possibly the
most natural initial condition in this context.
(Furthermore, we should note that the requirement that some
coupling constant is small is quite common even in standard
implementations of the inflationary scenario. For instance in chaotic
inflation with $\lambda\phi^4$ theory we must have $\lambda\sim
10^{-15}$ in order to obtain a correct value for the density fluctuations.)

In conclusion, there is a region of the parameter space of the model
that gives a long inflationary phase at the string
scale, while at the same time
the existing observational bounds on the production of 
relic gravitational waves are respected. 
This DeSitter inflationary phase
can be long enough to solve the horizon/flatness problems, or it can be
combined with the superinflationary phase to provide the 
required amount of inflation. In
the former case, the  value of the 
intensity of the relic gravitational wave spectrum to
be expected at ground based interferometers is of order
$\Omega_{\rm GW}\sim$  a few $\times 10^{-9}$ while in the latter
case it can reach a maximum value 
$\Omega_{\rm GW}\sim$  a few $\times 10^{-7}$.

\end{document}